\documentclass[%
 reprint,
 amsmath,amssymb,
 aps,
]{revtex4-2}

\usepackage{graphicx}
\usepackage{dcolumn}
\usepackage{bm}
\usepackage{float}

\begin{document}


\title{More than technical support: the professional contexts of physics instructional labs}

\author{LM Dana$^1$}
 \email{ldana@wpi.edu, she/her}
\author{Benjamin Pollard$^1$}
 \email{bpollard@wpi.edu, he/him}
\author{Sara Mueller$^2$}
\email{sara_mueller@brown.edu, she/they}

 \affiliation{$^1$Worcester Polytechnic Institute, Worcester, MA 01609, USA}
 \affiliation{$^2$Brown University, Providence, RI 02912, USA}

\date{\today}

\begin{abstract}
Most, if not all, physics undergraduate degree programs include instructional lab experiences.
Physics lab instructors, both faculty and staff, are instrumental to student learning in instructional physics labs.
However, the faculty-staff dichotomy belies the complex, varied, and multifaceted landscape of positions that lab instructors hold in the fabrics of physics departments.
Here we present the results of a mixed methods study of the people who teach instructional labs and their professional contexts.
Recruiting physics lab instructors across the US, we collected 84 survey responses and conducted 12 in-depth interviews about their job characteristics, professional identities, resources, and experiences. 
Our investigation reveals that lab instructors vary in terms of their official titles, job descriptions, formal duties, personal agency, and access to resources. 
We also identified common themes around the value of instructional labs, mismatched job descriptions, and a broad set of necessary skills and expertise.
Our results suggest that instructors often occupy overlapping roles that fall in between more canonical jobs in physics departments.
By understanding the professional contexts of physics lab instructors, the rest of the physics community can better promote and engage with their critical work, improving laboratory learning both for students and for the lab instructors who teach and support them.

\end{abstract}

\maketitle

\section{\label{sec:intro}Introduction}

Instructional labs are an essential part of formal physics instruction. 
They offer opportunities for learning distinct from theory courses \cite{AAPT2014}, and support such learning when they are structured around lab-specific learning goals \cite{Sulaiman2023,Walsh2022,Smith2021,Kalender2021, Smith2020c, Pollard2020a, Etkina2010}.
In parallel to such research on particular instructional lab structures and pedagogies, work has been done to investigate more fine-graned or elemental aspects of learning in physics labs. 
For example, researchers have studied processes and experiences specific to learning experimental physics \cite{May2022, Rios2019, Cai2021,Walsh2019a} and social dynamics and equity in lab group work \cite{Doucette2020, Stump2023}.
However, when it comes to the people who teach and support instructional physics labs, relatively little has been studied.
Research involving lab instructors has focused on documenting their perspectives on particular learning outcomes such as modeling \cite{Dounas-Frazer2018c} or measurement uncertainty \cite{Pollard2021a}.

It is important to understand the people who teach, support, and run physics teaching labs and the resources available to them. 
One reason for this importance is that the administration of lab courses tends to require greater infrastructure needs than theory-based courses \cite{Caballero2018}. 
At the introductory level, the labs may be coordinated with a theory lecture, fully integrated with theory components, or stand alone as their own course. 
In any case, intro labs are typically necessary to satisfy accreditation purposes. 
In labs beyond the first year (BFY), while such courses are typically stand-alone, the activities might be project-based, group-oriented, and/or writing focused. 
In each of these cases, the needs for those courses are often provided not just by the instructor of record, but also by additional staff or faculty supporters.
This wide variation in course structures, roles, and titles creates a complex and varied landscape for physics instructional lab professionals.

To the best of our knowledge, there is no previous work investigating physics lab instructors and support staff as a social or professional group.
Here, we seek to understand and document the structure, environment, and capabilities of the instructional support staff \footnote{We use the word ``staff'' to refer not just to people whose jobs are not considered faculty, but more generally to refer to anyone working to teach or support instructional labs.} that run instructional physics labs across North America. 
People in these positions often fall outside traditional academic roles in their university. 
They can be at very different places in their careers, and their positions can afford varying levels of access to resources and institutional power. 
This study of instructional lab personnel illuminates both issues and triumphs surrounding access to concrete resources (like equipment, space, and funding) as well as to human resources (such as personnel, student employees, and departmental support). 
We aim to answer the following research questions:
\begin{enumerate}
    \item In terms of job title, experience, and social identity, who are the people engaged in supporting and teaching instructional physics labs?
    \item How do they describe their roles, particularly with respect to their duties and agency?
    \item How do they perceive the value of instructional labs, both in their local context and more broadly by the physics community?
\end{enumerate}
Additionally, to add necessary context to the answers to these questions, we explore the following:
\begin{enumerate}
    \item What resources both in equipment and personnel are available to these folks?
    \item How does the availability of resources vary by institutional context?
\end{enumerate} 

Through answering these research questions, we aim to elevate, acknowledge, and support the people who 
work in physics instructional labs.
Understanding the professional context of such physicists is relevant to any lab transformation project, no matter its particular pedagogy or learning objectives.
Thus, beyond documenting for the sake of research, we hope this work will also be used to inform and improve future pedagogical reforms of physics lab courses.

\section{\label{sec:background}Background}
There has been much scholarship concerning the culture of physics overall and on sub-disciplines and sub-cultures within physics. 
Researchers have also studied the experiences of physicists and physics learners with particular social identities, especially those subject to marginalization and oppression.
We begin by briefly outlining that previous work in Sec. \ref{sec:physCulture}. 
Then in Sec. \ref{sec:higherEd}, we provide a brief overview of more general research that studies the structures and institutions of higher education in the US, focusing on staff positions but not specific to physics.
We do not claim for these overviews to be comprehensive; we intend only to point to the broader context in which our study is situated, highlighting the aspects that connect to the results of our work.
In Sec. \ref{sec:societies} we describe professional societies that connect and serve physics lab instructors, and then conclude in Sec. \ref{sec:positionality} with positionality statements that situate ourselves in these larger contexts.

\subsection{Physics culture(s) \label{sec:physCulture}}
Physics as a culture has been studied in a variety of ways, typically focusing on certain cultures within physics as a discipline.
One influential and established work on physics culture is \textit{Beamtimes and Lifetimes} by S Traweek \cite{traweek_beamtimes_1988}.
Traweek's work focused on the international culture of particle physics in the 1980s.
A notable aspect of physics culture in that work, and one that underlines much of the study presented here, is ``physics as a calling.'' 
Traweek writes (p. 21), ``The senior physicists who manage the lab are said to have been outraged at the very idea of a union, which suggested that people could consider their work at the lab as a mere job. The physicists generally have been committed to being scientist since early adolescence, and their own training teaches them to consider physics as a calling, not an occupation.''
The dichotomy between physics as a calling and as an occupation is not specific to the culture of 1980s particle physics.
It complicates the idea that physicists hold jobs purely for materialistic motivations.
Instead, Traweek suggests that working as a physicist is as much an aspect of identity as it is of income.

A second aspect of physics culture that Traweek identified is that physics is a ``culture of no culture.'' 
This aspect might seem contradictory to a culture that considers its work a calling rather than employment. 
However, more recent work by A. Hodari \textit{et al.} \cite{Hodari2022,hodari_policing_nodate} reconciles this dilemma though the notion of objectivity.
The authors of that paper write about how objectivity is central to (white) physicists' conceptions of themselves, and how this pervading culture of supposed objectivity often conceals underlying lenses and biases stemming from social, cultural, and institutional systems.
They identify the culture of no culture as a manifestation of ``the lie'' that, through a ``special form of mental gymnastics,'' allows physicists to embody such contradictions.
This phenomenon is especially salient when it comes to forms of identity-based oppression \cite{Hodari2022,hodari_policing_nodate}. 
Such complexity in physics culture overall is context for the professional and social identities of physics lab instructors; it is impossible to tell these stories without this context.

Other recent studies have focused on the crisis of culture in undergraduate physics departments with respect to the experiences and retention of undergraduate students.
In particular, women, gender minorities, and underrepresented racial/ethnic minorities in physics are a focus of active research. 
For example, the Team-Up report \cite{teamup}, a 2020 American Institute of Physics report that focused on implementing cultural changes for Black undergraduate students, presents a detailed description on how to help students cultivate a ``physics identity,'' indicating that physics is more than just a degree to find a job, but that one is entering a prestige community by becoming a physicist. 
A related body of scholarship examines the underrepresentation of women and other gender minorities in physics.
For example, Lewis \textit{et al.} \cite{lewis_fitting_2016} focuses on women in physics's sense of belonging, as a proxy for work that has highlighted that women leave physics because they did not feel a sense of belonging in the culture of physics \cite{CorrellShelleyJ.1997TaLW}.

\subsection{Staff in academia \label{sec:higherEd}}
While we are unaware of previous research on physics laboratory instructors and support staff, we draw upon a body of scholarship on staff (as distinct from faculty) in higher ed that is not specific to physics. 
By the numbers, staff in academia constitute almost half of all academic positions \cite{wilk_work-life_2016}.
Perhaps the most pervasive theme that emerges from this group as a whole is a picture of overwork, burnout and role ambiguity \cite{WinfieldJakeD2022BaWC,  mullen_job_2018, lorden_attrition_1998, wilk_work-life_2016}

In 1998 Lorden reviewed the literature up to that point on attrition in student affairs offices. 
At that time attrition rates ranged from 32\% in the first 5 years to 61\% within the first 6 years. 
They identified the reasons for leaving as limited opportunity for advancement and professional development, burnout, unclear job expectations, and conflicts between the values that motivated them to enter the field, the realities of their job, and low pay \cite{lorden_attrition_1998}. 
Twenty years later, in 2018, Mullen et al. surveyed 789 student affairs professionals job stress, burnout, job satisfaction and turnover intention, before the 2020 pandemic. 
They found that 21\% of the respondents reported moderate to high symptoms of burnout symptoms, and that burnout was predictive of respondents intention to leave the field \cite{mullen_job_2018}. 

The nature of staff jobs contributes to these phenomena. 
In 2016 Wilk compiled a comprehensive analysis of the work-life balance and expectations for administrators in academia. 
In that work they define ``administrators'' as a university worker who is salaried, focusing on workers in the student affairs office and technology group. 
They found that, "Although part of the larger university, each division had its own culture, expectations, and accepted practices." \cite{wilk_work-life_2016}. 
Their work identified a cultural expectation that held across groups: a pressure to work longer hours than contracted, to the point that 79\% of student affairs workers had difficulty finding a work-life balance. 
Wilk pointed to expectations by both supervisors and students, who expected 24/7 connection.  
Since 2020, research on academic staff has taken on even more dire tones, focusing on burnout in all roles.
For example, Winfred and Paris (2022) conducted a study surveying American collegiate registrars about their experiences and institutional policies (1,080 respondents), and their research found increased burnout, where the common cause was unsustainable workloads and an expectation of constant availability \cite{WinfieldJakeD2022BaWC}.



\subsection{Professional communities of physics lab instructors \label{sec:societies}}
There exist several professional organizations that include physics laboratory instructors, but only one whose mission is explicitly physics lab instruction focused. 
The Advanced Laboratory Physics Association's (ALPhA) mission is to provide communication and interaction among advanced laboratory physics instructors worldwide. 
Since 2007, ALPhA has served a specific subset of the community of physics lab instructors, namely, laboratory professions focused on instructional labs beyond the first year of undergraduate education. 
According to ALPhA, as of April 2023, they have over 200 members \footnote{https://advlab.org/about}. 
This constitutes a significant part of the advanced physics lab community: for context, as of 2021 there 754 universities in the United States that offer physics degrees at all levels, 494 of which offer Bachelors as their highest physics degree \cite{AIP_Undergrads}.
While multiple professionals from the same university are often members of ALPhA, it is not always assured that there is a laboratory specific professor or staff member at all of these universities.
In our view, ALPhA is the most substantially representative professional organization that is specific to physics lab instructors and supporters in the US.

In addition to ALPhA, there is an overlapping physics lab community within a larger professional organization, the American Association of Physics Teachers (AAPT). 
AAPT maintains several committees specifically on labs and equipment, and attracts a wider audience as it includes both the first year and BFY teaching lab community. 

Both AAPT and ALPhA hold summer conferences at regular intervals, offer workshops for professional development and idea sharing, and maintain email lists for asynchronous communication.
They act not only as supporting institutions for physics instructional labs, but also as hubs for networking and community building.

\subsection{Positionality statements \label{sec:positionality}}
We, the three authors of this work, wish to acknowledge and make explicit the positions, stances, and perspectives we bring to this study.
We do this as a best practice in social science research \cite{HolmesPositionality2020, Hampton2021} in order to give important context to our findings.
We drew on our positions and lived experiences when conducting this study, in particular during the initial visioning and planning phases, creating our interview protocol, and recruiting participants.  
During data analysis and interpretation we have worked to preserve and faithfully represent the perspectives of our interviewees, however, we acknowledge that in the end our own perspectives are inextricable from the analysis process.  

L.D. is the physics laboratory manager at Worcester Polytechnic Institute. She is both a graduate student working on her PhD, and a staff member who runs the introductory teaching physics labs at WPI. Her job duties involve writing lab material, maintaining equipment, managing the graduate and undergraduate students who teach the laboratories, teaching, and grading. Previously she was a technical instructor at MIT in the 8.13/8.14 lab course.

B.P. is a junior non-tenure-track teaching faculty member who regularly teaches the theory parts of large intro physics courses, and also teaches an upper-division lab course for physics majors. 
He mentors undergraduate and graduate students, including L.D., in Physics Education Research (PER) at his institution. 
He also works with L.D. in her capacity as the instructional lab manager for their department. 
His work in PER has focused almost exclusively on lab courses; it was through that work that he became familiar with the physics lab instructor community as a PER postdoc.
Previously, he completed his PhD as an experimentalist in the field of nano-optics. 
Though he considers himself a novice physics lab instructor and a ``labs person,'' he still orients to the lab instructor community primarily as a researcher and a relative outsider.

S.M. is a technical instructional labs staff and adjunct faculty member who supports the beyond first year laboratories, manages the instructional  staff, and teaches the graduate level laboratory course at their institution.
They completed their Ph.D. in experimental condensed matter physics and spent some of their graduate work using quantitative educational psychology tools like motivation to study physics education in graduate student populations.
They consider themselves to be a newcomer but insider to the laboratory instruction community.

While not a primary site of analysis of this work, social identities came up in our data and are discussed in our analysis below.
Moreover, social identities are relevant in social science research even if they are not explicitly studied.
Therefore we wish to acknowledge for additional context to our work that all of the authors of this study identify as early-career, white, and queer, with a range of gender identities, (dis)abilities, and socio-economic upbringings.
We note that our own social identities are generally well-represented in our study population as shown in Table \ref{tab:demographics}.
However, most of the study participants have been involved in instructional labs for longer than the authors.

\section{\label{sec:methods}Methods}
The mixed-methods approach we used for this study began with a 37 question survey at the end of which we solicited participation for follow-up interviews. 
We used the survey responses to get a broad overview of research questions (RQs) 1, 3 as well as contextual questions 1 and 2 (see Section \ref{sec:intro}). 
Our interviews formed a rich complement to the survey responses, allowing us to more deeply engage with RQ 2 and contextual question 1 in particular. 

\subsection{\label{methods:quant} Survey Methods}
In Spring 2022, the authors created a survey to collect information about the structure, environment, and capabilities of the instructional support staff and faculty who run physics laboratories across North America.
The survey contained a mix of free-response, multiple answer, and single answer questions exploring the respondent's job title, responsibilities, professional relationships (both locally and broadly), equipment and resource availability, perception of value, and demographic information.
The survey items are included as an appendix to this paper.
At the end of the survey, respondents were offered an opportunity to be interviewed for the study.
None of the survey questions were forced-response. 
While most of the participants answered each question, it may be the case that the total number of responses to an individual question is less than the total number of respondents. 
We initially solicited survey participation in-person at the AAPT 2022 Summer Meeting and then later through the ALPhA email list-serv.
In the first 45 days of the survey, 82 initial responses were recorded. 

In an effort to engage with institutions that are historically and presently under-supported and resourced, we developed a database for additional outreach.
Student workers created a spreadsheet from the NASA Minority Serving Institutions Exchange with the Physics filter and combined that with the institutions Carnegie designation, Geographic Location, City, State, AIP Institution Name, Highest Physics Degree offered, Astro Program, 2020-21 First-Term Introductory Physics Course Enrollment. \cite{nasa_msi, aipData, carnegie} 
From this database we created a randomized list of Native American-serving Non-Tribal Institutions, Historically Black Colleges and Universities, Asian American and Native American Pacific Islander Serving Institutions, Hispanic-Serving Institutions. 
Seeking to have equal proportion of these 5 populations represented in our list, we identified 40 institutions to reach out to, and the student workers were able to find the emails and contact information of 31 of them online. 
We then emailed those 31 universities with a link to our survey. 
We received three additional responses to our survey were from these institutions, and included them in our data set.
Information regarding the institutional characteristics of participants in this study are given in Table \ref{tab:institutions}.

Our survey was built and administered through Qualtrics \footnote{https://www.qualtrics.com/}. 
Likert response questions were analyzed with SPSS \footnote{https://www.ibm.com/products/spss-statistics} and open response questions were analyzed in MS Excel \footnote{https://www.microsoft.com/en-us/microsoft-365/excel}. 


\subsection{\label{methods:qual} Interview Methods}
We conducted semi-structured follow-up interviews with survey participants who opted-in to participating in such an interview.
Our interview protocol was developed from our initial RQs. 
Each question constituted a separate section, namely, ``Career path and organizational structure,'' ``Relationships and identities,'' ``Resources and agency,'' and ``Value from institution.'' 
Our protocol may be found in an appendix to this paper. 
As the interviews were semi-structured, follow up questions were utilized as time and interviewee's schedule allowed. 
Where interview subjects were inclined to expand they were encouraged to do so, with limited direction from the interviewer.

Many survey respondents from our initial round of outreach indicated that they were interested in doing a follow-up interview with us.
L.D. contacted these respondents via email and conducted the interviews.
Surveys were conducted remotely using video conference software \footnote{https://zoom.us/}. 
They ranged anywhere from 37 to 96 minutes.
There were a total of 12 interviews collected, occurring between August 5, 2022 and September 9, 2022.
Of those 12 interviews, only 11 were analyzed due to extenuating circumstances.
Recordings were automatically transcribed by OtterAI \footnote{https://otter.ai/}, and then initially reviewed by L.D. for correctness.

\subsubsection{Coding Process}
We conducted a qualitative analysis of the interview transcriptions consisting of two passes through our data set.
First, we performed a content analysis via deductive coding that was based on our research questions and interview protocol.
Second, we identified a subset of that initial content analysis and performed a thematic analysis via inductive coding to reveal emergent themes. 
We used both the NVivo \footnote{https://lumivero.com/products/nvivo/} and Dedoose \footnote{https://www.dedoose.com/} software applications to conduct our analysis.

We started with an \textit{a priori} set of codes based on our research questions, which also aligned with the questions in our interview protocol.
These codes were organized under two overall categories, ``Job characteristics'' of the interviewee's position and ``Resources'' available to the interviewee \footnote{We created additional categories and codes, but chose to focus on these two for this study. Additional analysis of our transcripts will perhaps be a focus of future work.}.
After creating this initial codebook, all three authors reviewed the first 10 or so minutes of one of the interviews synchronously and assigned codes to excerpts collaboratively.
In so doing, we shuffled, refined, and clarified our code definitions as needed.
Then, L.D. coded the remainder of that interview, applying codes to all excerpts she identified as relevant.
B.P. independently coded the excerpts identified by L.D. without looking at her coding. 
These two coding assignments were compared for inter-rater reliability (IRR).
We used Cohen's Kappa \cite{Cohen1960} as a measure of IRR, which yielded a kappa value (averaged across all codes) of 0.41 for ``Job characteristics'' and 0.87 for ``Resources.''
All three researchers met to discuss and reconcile all the ``Job Characteristics'' excerpts that L.D. and B.P. coded differently, further refining our code definitions as needed.
With this refined codebook, S.M. independently coded the entirety of another interview in our data set, and then B.P. independently assigned codes to the excerpts they had identified.
This second round of IRR yielded kappa values of 0.77 for ``Job characteristics'' and 0.85 for ``Resources,'' which we deemed a sufficient level of agreement to proceed.
L.D. and B.P. then coded the remaining interviews independently.

After this initial content analysis, we selected three of the \textit{a priori} codes for further thematic analysis.
We selected these codes based on both their prevalence in the data set and their relevance to our RQs.
Each researcher independently reviewed all of the excerpts labeled by one of these codes and created an emergent codebook to represent the themes they identified, assigning each excerpt at least one emergent code.
We subsequently met and collaboratively reviewed each others' thematic analyses, coming to consensus on our final emergent code definitions and applications after discussion.

\subsubsection{Member Checking}
After we synthesized our findings, created an outline of this manuscript, and identified the quotes we intended to present below, we performed a basic member checking process.
L.D. emailed each interviewee whose quotes we had selected, sending them our transcription of their quote and a brief explanation of our interpretation of it.
We asked them in our message if they were comfortable with the quote being used anonymously in this paper, and if the characterization of the quote was accurate.
They were given the option to respond via email or to schedule a Zoom call with L.D.
In total, four interviewees received such an email, and all four responded via email.
One agreed with no further feedback. 
The remaining three had requests to clean up the language of their quote while preserving the original meaning, and one also asked to clarify the meaning of our interpretation. 
After we responded to each via email, the remaining three interviewees agreed with our use of their quote and our interpretation of it. 
The final, agreed-upon language and interpretation are the quotes presented in this manuscript.

\section{Survey results\label{surveyResults}}
We present results from our survey responses across four main areas of interest for lab instructional staff.
We start by outlining the basic characteristics of the people who teach and support instructional labs, including their social identities and institutional contexts. 
We then turn to characterizing the resources, defined broadly, available for their work.
We distinguish these resources into two categories, first describing more concrete resources such as laboratory space and equipment, and second describing less tangible resources such as informational resources and social networks.
Then we describe the responsibilities of instructional lab staff, and conclude by presenting responses about the value of instructional labs.

\subsection{Characteristics of Lab Personnel}
\begin{table}[t]
\caption{\label{tab:demographics}%
The data set consists of 84 total survey respondents and 11 total interviewees. The entries in this table correspond to the percentage of respondents and number of interviewees that indicated the given characteristics in the survey. Per our data management plan, we do not show percentages less than 5\% to preserve the anonymity of our survey takers. 
}
\begin{ruledtabular}
\begin{tabular}{lcc}
\textrm{Characteristic}&
\textrm{Survey respondents}&
\textrm{Interviewees}\\
\hline \hline
Gender \\
\hspace{2mm} Women & 26\% & 3 \\
\hspace{2mm} Men & 68\% & 8 \\
\hspace{2mm} Non-binary & $<$5\% & 0 \\
\hspace{2mm} Prefer not to answer & $<$5\% &  0 \\
\colrule
\textrm{Education Level}&
\textrm{}&
\textrm{}\\
\hspace{2mm} Some college or Associates& $<5$\% & 0\\
\hspace{2mm} Bachelor's & 12\% & 2\\
\hspace{2mm} Master's & 17\% & 1\\
\hspace{2mm} Doctoral & 68\% & 8\\
\colrule
\textrm{Race or ethnicity}&
\textrm{}&
\textrm{}\\
\hspace{2mm} White & 94\% & 11\\
\hspace{2mm} Any other race\\ \hspace{4mm}and ethnicity & $<$5\% & 1\\
\colrule
\textrm{Role}&
\textrm{}&
\textrm{}\\
\hspace{2mm} Faculty & 52\% & 6\\
\hspace{2mm} Staff   & 48\% & 5\\

\end{tabular}
\end{ruledtabular}
\end{table}
The demographics of all respondents and interviewees are shown in Table \ref{tab:demographics}.
The survey questions contained more options than are presented in the table, in particular with respect to race or ethnicity, but we present aggregated categories here in order to preserve anonymity.
We note that an overwhelming number of our survey respondents identify as white, and that even less racial/ethnic diversity seems to exist in this population than in physics as a discipline overall. 
With regards to gender, 26\% of our survey respondents identify as women, which is slightly higher than the proportion in physics degree holders overall \cite{AIP_Women}.

From the survey results, two primary university roles of our participants were present: faculty and staff appointments. 
Using responses from the question ``What is your job title?'' we were able to label those who said ``Professor'', ``Lecturer'', and ``Instructor'' as faculty. 
All other job titles were designated as staff roles, and included titles like ``Laboratory Manager'', ``Instructional Technician'', ``Director of Undergraduate Laboratories'', and ``Lab Coordinator''. 
In the 84 respondents, we classified 44 in faculty roles and 40 in staff roles. 

\begin{table}[t]
\caption{\label{tab:institutions}%
Characteristics of the institutions with which our survey and interview respondents are associated, again given in percentage of survey takers and number of interviewees. 
All classifications were derived from the most recently available Carnegie database. MSI designations for campuses are HBCU: Historically Black Colleges and Universities, AANAPISI: Asian American, Native American, and Pacific Islander Serving Institutions, HSI: Hispanic Serving Institutions. One campus can have multiple designations.
}
\begin{ruledtabular}
\begin{tabular}{lcc}
\textrm{Institution Characteristic}&
\textrm{\% Survey respondents}&
\textrm{Interviewees}\\
\hline \hline
Highest degree granted \\
\hspace{2mm} Doctoral (R1)        & 41    & 4 \\
\hspace{2mm} Doctoral (R2, D/PU)  & 14    & 1 \\
\hspace{2mm} Master's only        & 12    & 6 \\
\hspace{2mm} Bachelor's only      & 25    & 0 \\
\colrule
\textrm{MSI Designation}&
\textrm{}&
 \textrm{}\\
\hspace{2mm} HBCU & 5\% & 0\\
\hspace{2mm} AANAPISI & 11\% & 1\\
\hspace{2mm} HSI & 11\% & 3\\
\hspace{2mm} None & 71\% & 9\\
\end{tabular}
\end{ruledtabular}
\end{table}

The primary differences we observed in the data from those in staff and faculty roles are in their stated responsibilities in the instructional labs and the distribution of their respective highest education levels.
However, their perceptions of how instructional labs are valued were not different based on their role. 

\begin{table}[h]
\caption{\label{tab:duration}%
Duration of involvement in physics instructional labs by role, in numbers of survey respondents.}
\begin{ruledtabular}
\begin{tabular}{lcdr}
\textrm{Duration in Instructional Labs}&
\textrm{Faculty}&
\textrm{Staff}\\
\colrule
$<$1 yr & 0 & 2\\
1 - 5 yr & 3 & 7 \\
5 - 10 yr & 13 & 9 \\
$>$ 10 yr & 28 & 20 \\
\end{tabular}
\end{ruledtabular}
\end{table}

Perhaps unsurprisingly, all those in faculty roles are doctoral degree holders.
In the staff designation, we observe that 37.5\% of the staff hold doctoral degrees, 30\% have master's degrees, 25\% have bachelor's degrees, and 7.5\% have some college or associate's degrees. 
Faculty and staff present a similar distribution of their tenure in instructional labs with both populations.
Many of the study participants indicate they have been working in the discipline for more than ten years, as shown in Table \ref{tab:duration}.
We asked participants who they work with most outside of their institution. 
Though we anticipated many of our respondents to be isolated, only 35\% indicated ``No one'' or ``N/A'' to this prompt.

These statistics present a snapshot of people, institutions, and job titles of physics lab instructors and supporters.
We present a more detailed description of survey respondents' responsibilities and job characteristics in Section \ref{sec:responsibilities} below.

\subsection{Resources and equipment \label{sec:resourcesEqupiment}}

\begin{figure*}
\includegraphics[scale=0.33]{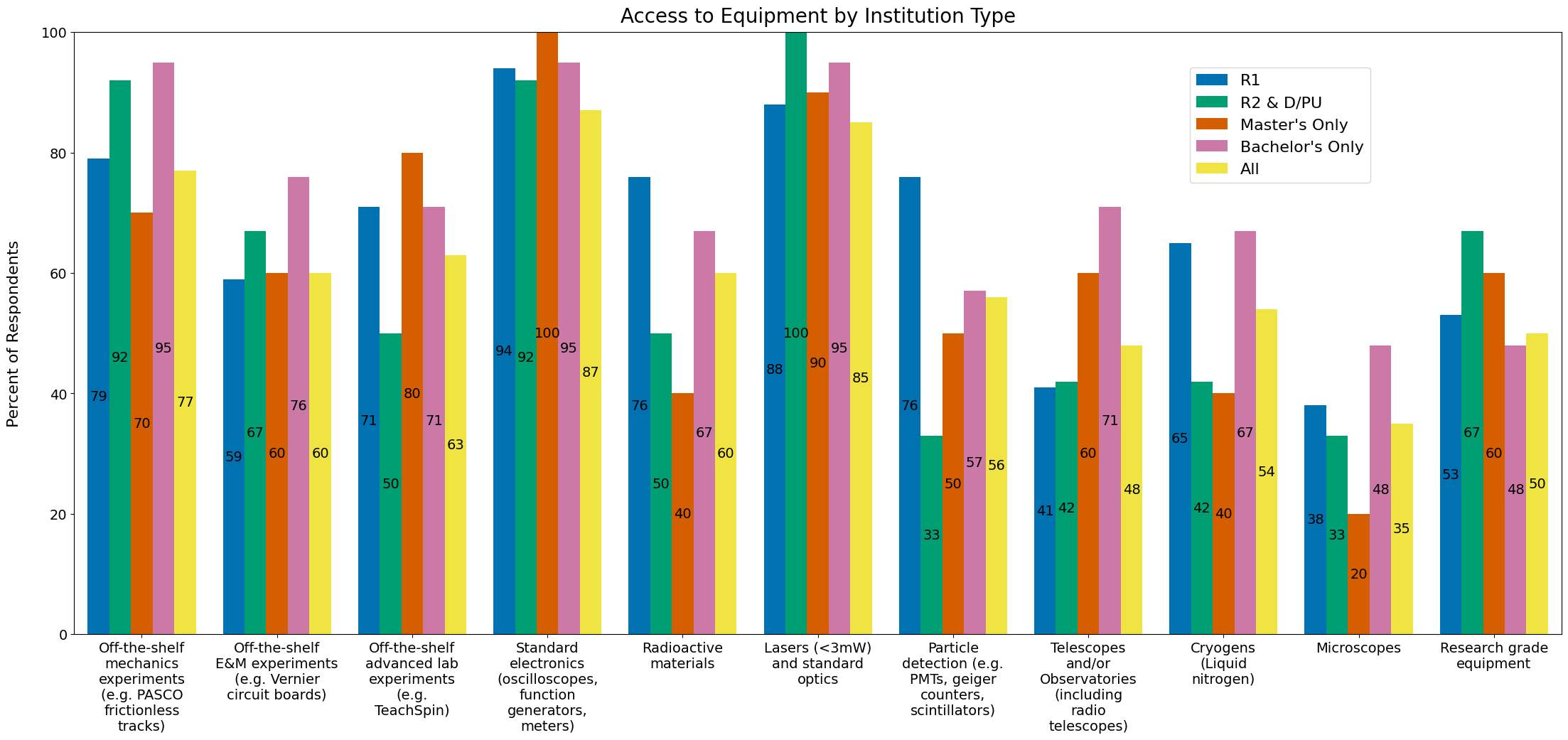}
\caption{\label{fig:equipment} Percentage of institution types with access to the equipment types along the horizontal axis.}
\end{figure*}

In our own roles in the labs, we know that access to resources and equipment can be a major factor in our success in the laboratory space. 
L.D. and S.M. developed a list of commonly used equipment in the instructional labs. 
Figure \ref{fig:equipment} shows the availability of this equipment disaggregated by characteristics of the institution from which the respondent comes.
We note that nearly all institutions have access to standard electronics and optics in their instructional labs. 

Off-the-shelf equipment, such as PASCO \footnote{https://www.pasco.com/} frictionless tracks, Vernier \footnote{https://www.vernier.com/} circuit boards, and TeachSpin \footnote{https://www.teachspin.com/} experiments, seem to be available to most of the respondents in this group. 
That is not to say that they are ubiquitous, however, that this type of equipment likely plays a crucial role in instructional labs.

In contrast, access to cryogens, microscopes, and research grade equipment is lower across all institution types. 
Prefacing the findings below (from Figure \ref{fig:responsibilities}) that most of the purchasing, repairing, and maintaining equipment falls to those in staff roles, and from the interview responses in Section \ref{orgStructureCodes}, that instructional labs staff are ``wearing many hats,'' it is not surprising that more complicated equipment and resources that require more maintenance are rarer.

We also note that though we might naively assume `R1' institutions to have more resources across the board, we observe that for many categories, including `Research grade equipment,' `R2 and D/PU', and `Master's only' institutions have a higher likelihood to have access to these resources.
We suggest that this is likely due to instructional labs pulling double-duty as both an instructional space and a research one in these contexts.

With respect to human resources, a majority of our survey respondents indicate that they interact with faculty, other staff, or student employees either `Daily' or `Weekly'.
More than half of our respondents indicate they are connected to other instructional labs staff at different institutions.
More insight into the types of resources available to instructional lab staff emerged from our interviews, as discussed below.

\subsection{Responsibilities and Job Characteristics \label{sec:responsibilities}}

\begin{figure*}
\includegraphics[scale=0.35]{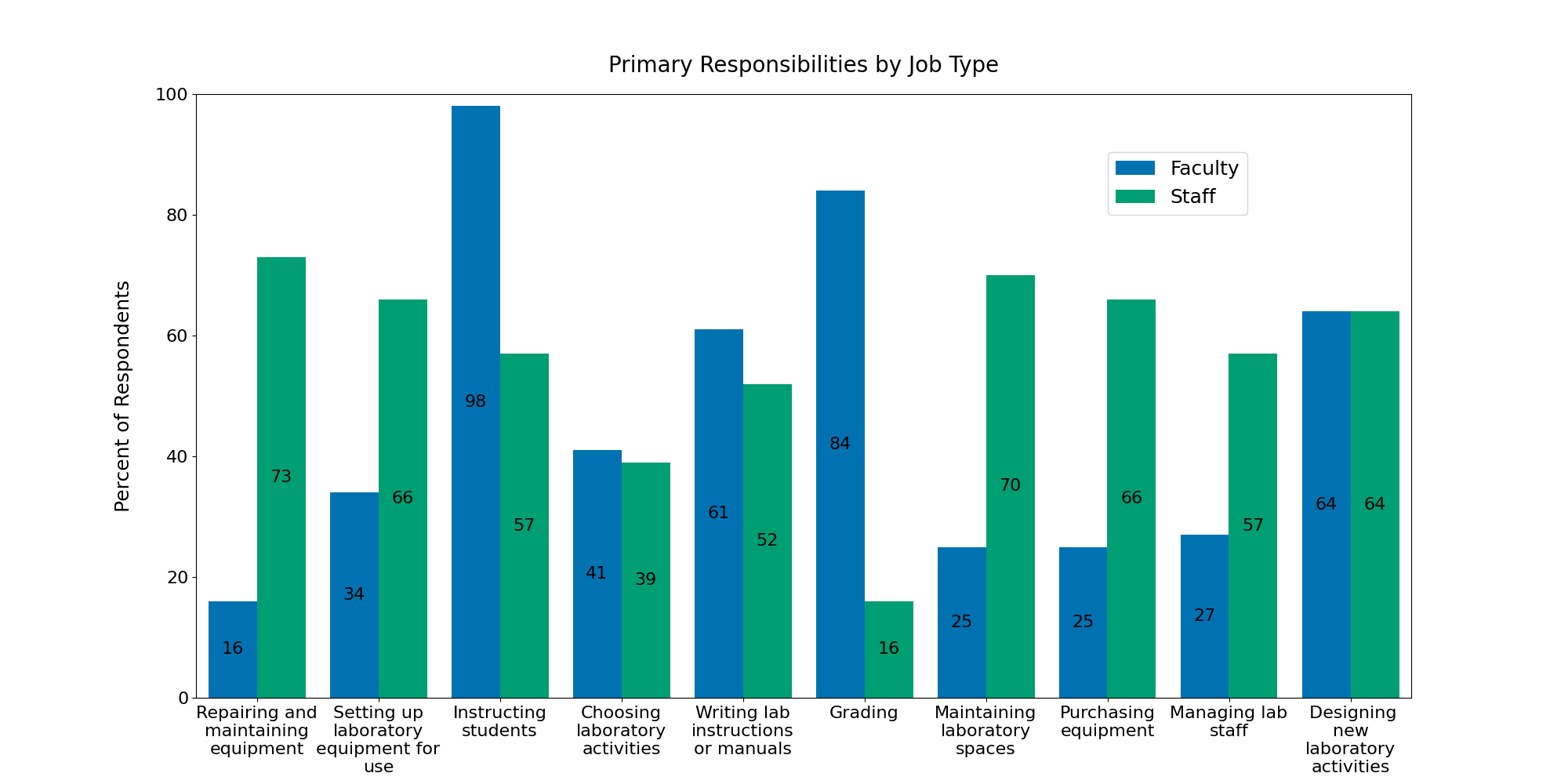}
\caption{\label{fig:responsibilities} Percent of faculty or staff respondents who selected each responsibility along the horizontal axis.}
\end{figure*}

We asked survey respondents to select any of the responsibilities in Figure \ref{fig:responsibilities} that require more than 20\% of their work hours during the course of the year.
The responsibilities we anticipated labs staff and faculty to hold ranged from ``Repairing and maintaining equipment'' to ``Grading'' and ``Instructing students'' to ``Designing new activities.''
We additionally added an ``another responsibility not listed'' option with text input. 
The only responsibility written in more than once was ``Safety officer,'' which indicates that the list of responsibilities given in the survey to be an overall robust description of the tasks associated with instructional labs staffing roles.

We notice two distinct but related arenas of responsibility: logistics and pedagogy.
For staff respondents, a higher proportion of their primary responsibilities fall into what we term ``logistical'' support for laboratories.
These responsibilities include ``Repairing and maintaining equipment'', ``Maintaining laboratory spaces'', ``Purchasing equipment'', ``Setting up laboratory equipment for use'', and ``Managing lab staff''.
In contrast, and as we naively would expect, the pedagogical tasks of the lab are more often primary responsibilities of faculty respondents.
For example, nearly all faculty respondents selected ``Instructing students'' as a primary responsibility in their role. 
There is, however, significant overlap in the two classes of job title, particularly in designing new laboratories and writing the instructions for their use.

These responsibilities are closely tied to the organizational structure in which lab instructors operate. 
We offer insights into these various organizational structures through our interviews, as discussed in Section \ref{sec:orgStructure} below.

\begin{figure}[h]
\includegraphics[scale=0.45]{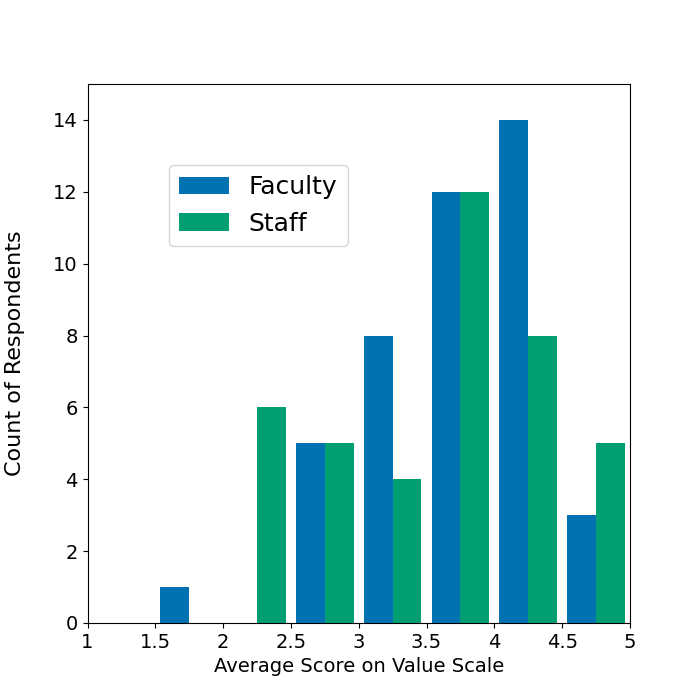}
\caption{\label{fig:value} Equivalent agreement statement based on the average score on the value scale from Table \ref{Value}.}
\end{figure}
\begin{figure}[h]
\includegraphics[scale=0.45]{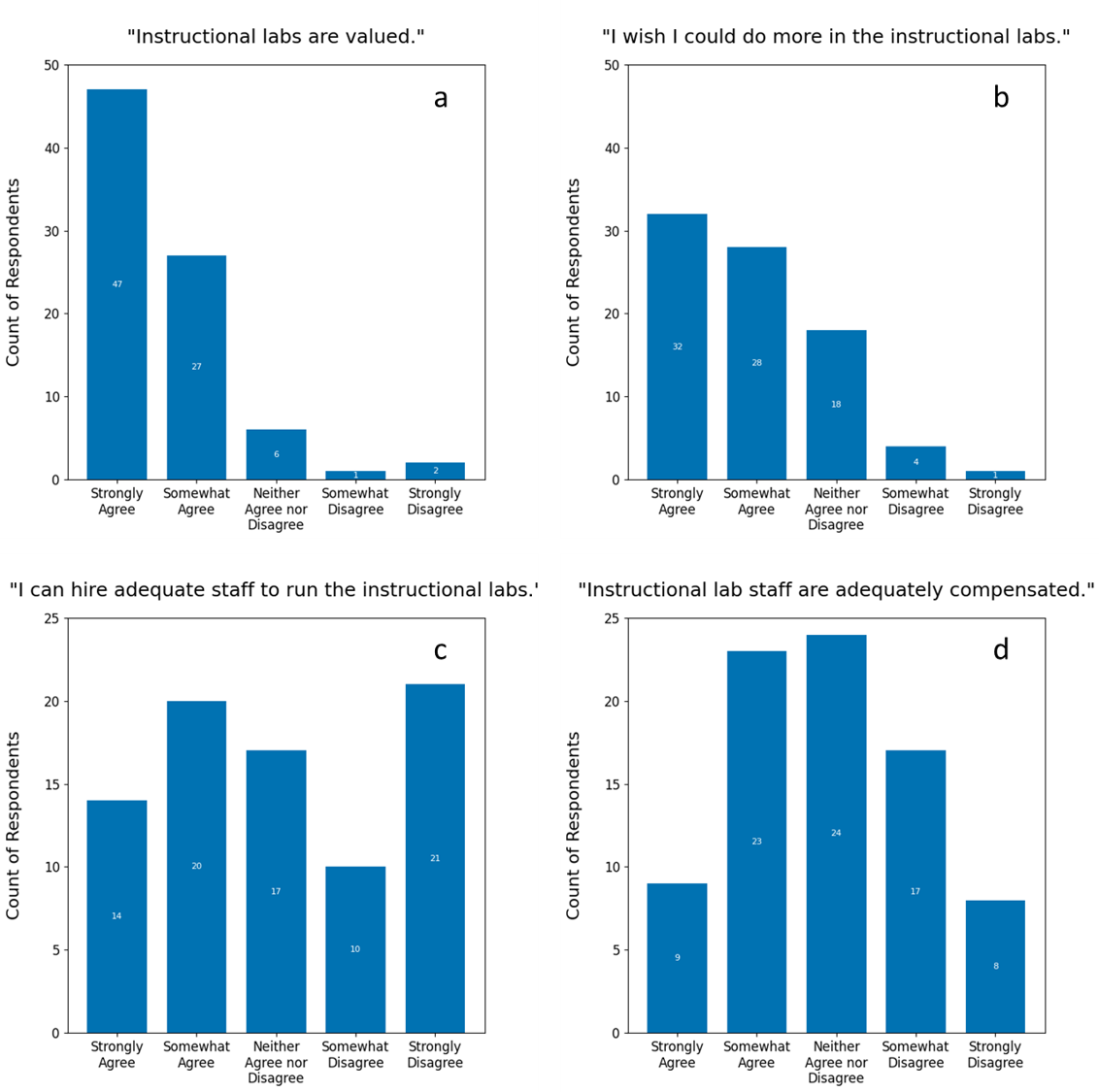}
\caption{\label{fig:value_items} Individual distributions on select items from the value scale from Table \ref{Value}. Across each row the y-axis has the same scale for direct comparison (panels a and b have a max is 50, panels c and d have a max is 25). a. ``Instructional labs are valued'', b. ``I wish I could do more in the instructional labs.'', c. ``I can hire adequate staff to run the instructional labs.'', and d. ``Instructional lab staff are adequately compensated.''}
\end{figure}

\subsection{Perceived value of instructional labs}
\begin{table*}[]
\caption{\label{Value}%
Instructional labs value scale.
All items were administered and answered in the survey. Items 7 and 9 are reverse coded in the analysis.
}
\begin{tabular}{p{.05\linewidth}|p{.66\linewidth}}
\textbf{}  & \textbf{Item} \\ 
\hline \hline
1 &  Instructional labs are valued. \\
2 &  Instructional lab staff are adequately compensated. \\
3 &  I have agency to make choices about the instructional labs. \\
4 &  I can hire adequate staff (students and professionals) to run the instructional labs. \\
5 &  I have sufficient funding to provide high quality instructional labs. \\
6 &  I am satisfied with the instructional labs at my institution. \\
7R &  I wish I could do more in the instructional labs. \\
8 &  I enjoy my career  in the instructional labs. \\
9R &  I would rather be working in another position at my university. \\
10 &  I am valued by the faculty in my department. \\

\end{tabular}
\end{table*}
As part of the survey, we developed a 10-item scale to examine the perceived value of instructional labs by those working in the field.
The items we used for this work are shown in Table \ref{Value}. 
All respondents answered these items on a 5-point Likert scale from strongly disagree with the statement to strongly agree with the statement.
Responses to this scale were well behaved with a Cronbach's alpha of 0.828, which corresponds to a good internal consistency between responses on items.
Given that the scale was sufficiently well behaved, we took an average score on all ten items for each respondent and plotted them disaggregated by role in Figure \ref{fig:value}.
The distribution of the faculty responses (blue) peaks at a score of 4, corresponding to a high perceived value of instructional labs, with a relatively long tail. 
The staff response distribution is flatter and peaks in the 3.5 bin.
However, by both ANOVA and Mann-Whitney testing there is no significant difference in the distributions.

Certain individual items on this scale showed notable behavior, as demonstrated in Figure \ref{fig:value_items}. 
Though answers to the item ``Instructional labs are valued'' indicate overwhelming agreement, responses were similarly high for the item ``I wish I could do more in the instructional labs.''
This combined with the middling responses to items regarding hiring and compensation (Figure \ref{fig:value_items}c, d) indicate that there is a desire in this community for more resources to do the high quality work expected of them.

\section{Interview results and discussion}
Our interview data set comprised transcripts with 11 interviewees.
Information about those interviewees and their institutions is shown in the rightmost columns of Tables \ref{tab:demographics} and \ref{tab:institutions}.
We provide that information for context of our interviewee sample population.

\subsection{Deductive content analysis}
\begin{table*}[t]
\caption{\label{firstPassCodes}%
The codes that resulted from our initial content analysis of the interview transcripts.
The \textbf{bold} headings denote the category of the codes below them.
The rightmost column counts the number of interviews (out of 11 total interviews) that contained at least one excerpt to which the code was applied.
}
\begin{tabular}{p{.25\linewidth}|p{.66\linewidth}|c}
\textbf{Code}  & \textbf{Description} & \textbf{Count}     \\ 
\hline \hline
\textbf{Job characteristics} &  \textit{Discussions about...} \\ \hline 
Organizational structure & ...the organizational structure of the labs or courses that the interviewee discussed. This code includes discussions about job responsibilities. & 10\\ \hline
Institutional hierarchy & ...how the interviewee's job or role fits into their institution's larger organizational structure of faculty, staff, and students. & 11\\ \hline
Career path & ...the interviewee's previous professional roles and titles, especially those that led them to their current position. & 11\\ \hline
Value of instructional labs & ...how and to what extent instructional labs are valued. & 11\\ \hline
Value of personal effort & ...how and to what extent the interviewee's work is valued. & 11\\ \hline
\textbf{Resources} &  \textit{Access to...} \\ \hline 
Informational resources & ...lab guides, manuals, instructional videos, and other written or electronic materials and documentation. & 6\\ \hline
Education and formal training & ...formal degree programs, continuing education programs, or official professional development opportunities. This codes includes both current opportunities and previous experiences. & 9\\ \hline
Equipment/apparatus/ money/space & ...physical devices and equipment for use in instructional labs, money devoted to instructional labs, or teaching space devoted to instructional labs. & 11\\ \hline
Local faculty and staff workers & ...the expertise and support of faculty and staff colleagues at the interviewee's current institution. & 11\\ \hline
Colleagues outside institution & ...the expertise and support of colleagues outside the interviewee's current institution. & 7 \\\hline
Personal agency & ...freedom to make pedagogical or logistical decisions about instructional labs, or about the interviewee's job structure. & 11\\ \hline
Personal knowledge and skill & ...particular skills, abilities, or expertise possessed by the interviewee. & 9\\ \hline
Student workers & ...students to help teach, prepare, grade, or otherwise support the job of the interviewee, or instructional labs in general. & 10\\ \hline
\end{tabular}
\end{table*}

Our first pass of coding analysis resulted in a set of codes representing the content of the interviews.
Those codes are shown in Table \ref{firstPassCodes}, along with accompanying descriptions and the number of interviewees who spoke about the code at least once during their interview.
These numbers show that all codes were well represented in our interviews, suggesting that instructors had substantial things to say concerning our research questions, and that the overall topics of inquiry resonated with them.

The codes and counts under the ``Resources'' category broadly align with the survey findings presented above (Section \ref{sec:resourcesEqupiment}).
This alignment suggests that the resources discussed in the interviews are representative of our larger survey respondent population.
However, certain codes under the ``Resources'' category had relatively fewer interview counts compared to the rest of the codes, specifically ``Informational resources'' and ``Colleagues outside institution.''
One might conclude that the lab instructors we interviewed lacked access to written instructional materials, and that they had few professional connections beyond the local level.
On the contrary, we wish to stress that the counts in Table \ref{firstPassCodes} under ``Resources'' are not an indication of whether interviewees had access to each resource, but rather how salient or important that resource was to the topics in our interview protocol.
We suggest that these counts represent not the availability of resources, but their relevance to lab instructors in their professional lives.
It is reasonable to assume that the things lab instructors chose to discuss are foremost on their minds, and therefore most impactful to their work.

In particular, we note that the lack of ``Colleagues outside institution'' counts aligns with the finding above (Section \ref{sec:resourcesEqupiment}) that the majority of survey respondents are connected to other instructional labs staff at different institutions.
However, there is a potential bias in our data set towards individuals who are connected as such.
We discuss this limitation below (Section \ref{sec:limitations}).

\subsection{Inductive thematic analysis}
Our second pass of coding analysis focused on the excerpts of three codes from the first pass: ``Organizational structure,'' ``Personal agency,'' and ``Value of instructional labs.''
We discuss the emergent themes from each of those topics in turn.

\subsubsection{Organizational structure \label{sec:orgStructure}}
\begin{table}[h]
\caption{\label{orgStructureCodes}%
The codes that resulted from an inductive thematic analysis of the excerpts coded with Organizational structure.
These codes are organized into two categories: one relating to the particular responsibilities of the interviewee's job, and one relating to the overall characteristics of their job.}
\begin{tabular}{p{.4\linewidth}|p{.6\linewidth}}
\textbf{Code}  & \textbf{Description}    \\ 
\hline \hline
\textbf{Job responsibilities} & \textit{Interviewee is responsible for...} \\ \hline 
Build and fix equipment & ...repairing apparatus, or building new apparatus.\\ \hline
Develop & ...conceiving, developing, and refining new lab activities or entire lab courses.\\ \hline
Safety officer & ...ensuring appropriate safety procedures are followed, or that students and colleagues are sufficiently educated on matters of lab safety.\\ \hline
Train / supervise & ...offering training sessions for, or ensuring accountability of, student workers or other lab staff\\ \hline
\textbf{Job characteristics} &  \textit{Interviewee describes their job as...}\\ \hline 
Low effort job & ...requiring relatively little effort.\\ \hline
On call & ...needing to be present and ready to respond to unexpected or in-the-moment events or requests.\\ \hline
Only one & ...being the the only person who works with instructional labs, or a certain type of instructional lab. This code includes descriptions of the difficulty to train people to help out the interviewee. \\ \hline
Permanent & ...expected or assumed to be available to them in perpetuity, unless they do something egregious. This code includes descriptions of tenure, both as an official policy and as an unofficial understanding.  \\ \hline
Unofficial / unclear & ...not well defined or unclear, at least to some extent, or involving expectations that are vague or unwritten.\\ \hline
\end{tabular}
\end{table}

The themes that emerged under ``Organizational structure'' are represented by the codes shown in Table \ref{orgStructureCodes}.
These themes clustered under two categories. 
We found that interviewees described the structure of their courses and roles in terms of both the particular job responsibilities they are responsible for, and the overall characteristics of their work.
Here we focus on two themes that concern these overall characteristics, as (in our view) they illustrate the qualities of lab instructor work that are atypical, or at least most distinct from other roles in typical physics departments.

One theme is represented by the ``Unofficial / unclear'' code, illustrating that the roles of lab instructors are often vague, undefined, or not well specified. 
For example, a section of one of the interviews with a staff interviewee was transcribed as,
\begin{quote}
\textbf{L.D.:} What is your job description? In your own words? \\
\textbf{Interviewee:} Okay, in my own words, I couldn't do that. Because I don't know that I have an official job description...it's there somewhere, I'm sure. But I don't know that I've looked at it or anything. \\
\end{quote}
Immediately after this exchange, the interviewee continued to describe in detail their actual job activities and what they are in charge of in practice.
This passage illustrates that while lab instructor positions usually have official job descriptions, these descriptions are often disconnected from the actual work of instructors and tend to lose relevant meaning ``on the ground.''
We infer, and the interviewee later confirmed during member checking, that one reason for this disconnect could be that lab instructor jobs require high levels of technical skill.
That qualification comes with agency (see Table \ref{agencyCodes} and discussion below), but also involves a fair amount of tasks that are tedious or mundane, and could even be seen as drudgery.
However, because of the hierarchy in which these jobs exist and the technical skill required, such tasks often cannot be delegated, and therefore remain stuck with the lab instructor.
As such responsibilities accumulate, the job becomes more and more disconnected from its official job description.

This idea was further indicated in our survey results.
In the open response ``other'' option from Figure \ref{fig:responsibilities}, several survey respondents wrote: ``Not at 20\% level, but I do all of these.'' 

The other theme we highlight here is represented by the ``Only one'' code.
This code describes being the only person doing the particular work of the interviewee.
While far from ubiquitous in our data set, a subset of interviewees spoke about the implications of being the only person at their institution associated with instructional labs overall, or with certain labs in particular.
For example, at one point in the interview protocol interviewees were prompted to discuss their prior response to the survey question about their primary responsibilities, and one faculty interviewee said,
\begin{quote}
So I checked them all, and I have done [them all]. I do less of it now. But I did a lot of it out of, one, interest. And two, because I was the faculty member in charge of our advanced lab class for a while. And so there was basically no one else doing it, or was going to do it. And so I did it.
\end{quote}
Because the interviewee was the only one, or one of a very few, interested in physics instructional labs, they ended up doing a wide range of tasks and serving many roles.
As a general trend, we found that lab instructors tend to ``wear many hats'' that require a range of overlapping skills and expertise.
This trend is also exemplified by the survey results presented above in Figure \ref{fig:equipment}, which shows that lab instructors access and use a wide range of equipment for their jobs.
This wide range of equipment suggests that lab instructors are expected to posses significant and broad range of technical expertise, not to mention the additional non-technical skills necessary for their jobs.
We suggest that such a breadth of overlapping skills and expertise is distinctive of lab instructor positions, representing a combination of proficiency that is broader than many other roles in physics departments.
This phenomenon is one of several we observed that suggest that lab instructor roles tend to fill in the gaps between more canonical roles in higher ed (e.g. pedagogy, technical expertise, or logistics).
In that way, we suggest that instructional lab staff occupy a liminal space, in the sense of being \textit{in between} the jobs of their colleagues.
Much work on liminality exists in anthropology, sociology, and beyond, and we do not claim to fully embrace those connotations here.
Nonetheless, we suggest that elements of perpetual shifting and a sense of ambiguity characterize the experiences of many lab instructors and support staff. 

\subsubsection{Personal agency}
\begin{table}[b]
\caption{\label{agencyCodes}%
The codes that resulted from an inductive thematic analysis of the excerpts coded with Personal agency.}
\begin{tabular}{p{.25\linewidth}|p{.75\linewidth}}
\textbf{Code}  & \textbf{Description}    \\ 
\hline \hline
& \textit{Interviewee has (or does not have) the agency to...} \\ \hline
Fun & ...incorporate aspects of play or fun into their work, and to express those aspects freely.\\ \hline
Instructor vs supporter & ...act as the main person with responsibility in instructional labs. This code concerns how agency is shared between the interviewee and their colleagues. \\ \hline
Learning objectives  & ...determine the learning objectives of the instructional labs that they teach.\\ \hline
Flexibility  & ...determine their day-to-day schedule, how they spend their time at work, or how they structure their work on a practical or logistical level.\\ \hline
\end{tabular}
\end{table}

The themes that emerged under ``Personal agency'' are represented by the codes shown in Table \ref{agencyCodes}.
These themes concern the ways in which interviewees feel a sense of agency (or lack thereof) over their professional activities, which range from the type of role they perform in the instructional lab to less tangible aspects that give interviewees a sense of agency in their professional lives.
Here we focus on one particular theme that we feel is most distinctive to lab instructor roles when it comes to personal agency: ``Instructor vs supporter.'' 

The ``Instructor vs supporter'' theme represents a dichotomy between two types of roles that lab instructors perform.
Interviewees discussed their jobs falling at various locations on a spectrum between those two roles, and also moving between them on a day-to-day or even moment-to-moment basis.
For example, a section of one of the interviews with a staff interviewee was transcribed as,
\begin{quote}
\textbf{L.D.:} How much control do you have in determining the learning goals of your labs? \\
\textbf{Interviewee:} Ah, depends on the instructor. But most of them are pretty darn high. Like to 100\% or so. Some courses where it's just, ``Here you go, can you do the lab thing.'' And it's all me, and that's perfectly fine. Other professors, they want to take a more hands on approach, which I totally support, because they're your students more than they're mine. And so then they can decide how they want to run the labs.
\end{quote}
This response illustrates how even a single lab instructor can find themselves in either a lead or a supporting role depending on who they are working with that term.
It is notable that this interviewee was not describing the inconsistency of their role in a negative light; they were both fine with taking charge of the lab, while also supportive of their colleagues to decide how the labs are run.
Here again we see an example of lab instructors shifting between different roles, tying in with the liminality thread discussed above.
Interviewees described how instructional lab personnel must be able to navigate this \textit{in-between}-ness in order to be effective in their jobs. 

\subsubsection{Value of instructional labs \label{surveyValueResults}}
\begin{table}[t]
\caption{\label{valueOfLabsCodes}%
The codes that resulted from an inductive thematic analysis of the excerpts coded with Value of instructional labs. 
These codes are organized into two categories based on whether they describe labs being valued (Positive) or not valued (Negative).}
\begin{tabular}{p{.25\linewidth}|p{.75\linewidth}}
\textbf{Code}  & \textbf{Description}    \\ 
\hline \hline
\textbf{Positive} & \textit{Instructional labs are valued because...}\\ \hline
Awards and formal recognition & ...lab courses (or lab instruction more generally) have been formally recognized, for example by university awards or formal public commendation, or in tenure decisions and performance reviews.\\ \hline
Dedicated resources to support labs & ...they receive tangible resources from the department, institution, or individual colleagues. Such resources include funding, equipment, student support, protected time for the interviewee, and dedicated space for instructional labs.\\ \hline
Interest from colleagues & ...administrators or faculty show interest and support for instructional labs, for example by visiting the labs, talking them up, or participating in discussions about lab logistics or learning goals.  \\ \hline
Job security / permanence & ...the position of the interviewee is secure, which affords them a level of respect to advocate and express themselves. This security also allows colleagues to rely on them for their expertise and service to the department.\\ \hline
Place in curriculum & ...they are worth significant course credit, are scheduled intentionally, or have explicit descriptions or requirements in the course catalog. \\ \hline
Student learning outcomes & ...students learn valuable things in instructional labs, especially when that learning is backed up by formal or anecdotal evidence, including from students themselves, that the process of learning in labs is worthwhile.\\ \hline
\textbf{Negative} & \textit{Instructional labs are not valued because...}\\ \hline
Insufficient / mismatched support  & ...the interviewee is left to ``pull up the slack'' created by a lack of support for instructional labs, or support that is not aligned with what is needed. Lack of support can include insufficient or inappropriate space or too high student-to-instructor ratios. Misalignment can include finding extra work for the interviewee to do in order to justify their position.\\ \hline
Question labs' value  & ...colleagues, students, or the interviewee themself questions whether labs are worth the cost, or the stress, work, or time that labs demand of students.\\ \hline
\end{tabular}
\end{table}

The themes that emerged under ``Value of instructional labs'' are represented by the codes shown in Table \ref{valueOfLabsCodes}.
They represent the things lab instructors indicated as evidence that instructional labs were valued (or not valued).
While related to how instructors themselves were valued, these themes specifically concern the value of instructional labs rather than the interviewees themselves.
We present these themes in two groups: ``Positive'' themes describe ways that labs are valued, while ``Negative'' themes represent ways that labs are not valued.
Here we focus on one negative and three positive themes to illustrate potentially interesting aspects in how interviewees talked about the value of instructional labs.

One faculty interviewee spoke of how labs are valued by saying,
\begin{quote}
So particularly before we started this curriculum rewrite, the labs were seen as something that we had to do for accreditation purposes, but nobody really cared about them. Before I arrived at [my institution] many courses used this very strange to me format where you know, lab was worth 100 points towards your grade. And as long as you got 70\% or more, you got all 100 points. So you just had to show up to the lab and kind of do it. And then you got all the points. To me that's not valuing lab...When I came in, the course I taught, I said, we're not doing that. I said, you're getting a grade for lab, and that's your grade. Because I valued labs...And anyway, so I feel like the labs are a little more valued than they used to be. There's still this mix of some instructors [who] see it as something that they have to do and it doesn't have much pedagogical value, others see it as more important.”
\end{quote}
We identified two themes in this quote: ``Place in curriculum'' and ``Question labs' value.''
With respect to the ``Place in curriculum'' theme, this quote describes an old and a new grading scheme as indicative of labs' value. 
The interviewee associates a separate and complete lab grade with labs being valued, and a scheme that amounts to ``complete/incomplete'' with not valuing labs.
The way the interviewee's lab is graded is tied to their lab's relationship to theory courses and the overall physics curriculum; by grading labs as a separate entity, they are valued as a distinct aspect of learning on par with other settings for physics instruction.
Valuing labs for their distinct learning potential parallels existing discourse in physics education research on labs and their learning goals, mirroring claims that labs have distinct learning goals from theory courses \cite{AAPT2014} and should be valued as such \cite{Caballero2018}.

The beginning and end of the quote highlights another theme in Table \ref{valueOfLabsCodes}, ``Question lab's value.''
At the start of the quote, the interviewee implies that they had agency to change the grading system because ``nobody really cared about'' the labs when they arrived.
Further reinforcing themes regarding personal agency discussed above, this lab instructor found agency in their professional context when they were left to make decisions by default.
However, the interviewee describes that, even after their grading changes, some of their colleagues recognize labs' pedagogical value, while others dismiss labs as unimportant pedagogically.
Such dismissal amounts to a tacit questioning of labs' pedagogical value, suggesting that grading schemes are but one piece of valuing instructional labs.

We provide a final example of labs being valued, this time via ``Student learning outcomes'' and ``Interest from colleagues.''
A faculty interviewee stated,
\begin{quote}
Companies come in and say, ``Oh, we love your students,'' and...``we need more students, because they can do this.''...One of my students, [name], he ended up getting a bunch of great jobs and ended up at [prestigious national lab], and then [that lab] said, ``How'd you get so good?'' He told them about us, and then they came and found us...We're small enough and make enough noise, and it seems like all the administrators through the President [at my institution] knows about us and knows that we turn out students that get good jobs and do good things. I mean, the President comes by our labs.
\end{quote}
This passage describes a specific, concrete example of student learning as evidence of labs' pedagogical value, and also points to administrators showing support for labs because of that learning.
Furthermore, the example illustrates the value of labs in providing students with transferable skills required to succeed in professional research, skills beyond what is learned in the standard theory curriculum.
Such an outcome is aligned with the larger discourse of researchers and policymakers around laboratory learning, career success, and workforce development \cite{Aiello2021,Fox2020,Leak2018,olson2012engage}. 
This interviewee is keenly aware of that larger context, and as their quote illustrates, witnesses its manifestation firsthand.
More generally, this quote is evidence that physics lab instructors are well positioned to act as powerful stakeholders in efforts to not only value labs in educational systems, but also to maximize that value through education reform and scholarship.

Overall, we heard from most lab instructor interviewees that their labs are valued as a whole, confirming what we learned from the quantitative results in Figure \ref{fig:value_items}a. 
However, as these themes show, the evidence our interviewees point to spans a wide range of aspects of professional life and laboratory learning.
Taken together, these results suggest that just as lab instructors often find themselves in between roles and wearing many hats, so too are their labs valued in a range of ways and contexts.
In fact, while we focus on the ``Student learning outcomes'' and ``Place in curriculum'' themes above, we wish to highlight that these are the only two themes in Table \ref{valueOfLabsCodes} that deal explicitly with pedagogy and student learning. 
The range of themes that emerged here illustrates that the value of instructional labs covers a much broader scope in the eyes of these lab instructors than, perhaps, the current scope of physics education research on instructional labs.
As such, we interpret that instructional labs themselves can exist in a liminal space between (and beyond) the current ontological systems of physics education research.

Additionally, the themes in Table \ref{valueOfLabsCodes} support and deepen the survey results presented above in Section \ref{surveyValueResults} on the value of instructional labs.
Our survey results show that lab instructors feel that their labs are valued overall, but their responses were mixed when asked about particular ways and aspects that value manifests.
In particular, the survey results show that when framing value in terms of the systems and bureaucratic structures in which labs exist (namely, hiring staff and compensation), the value of labs varied from very high to very low. 
As in the interviews, when our scope of analysis broadens from more traditional student learning outcomes, the picture of labs' value becomes more complex.
Taken together, we suggest that these constraints in resources can obscure or override conversations about pedagogy and learning.

\subsection{Scope and Limitations \label{sec:limitations}}
The primary limitation of this work stems from our methods of recruiting survey and interview participants.
Our recruitment approach relied on existing national professional networks, namely AAPT and ALPhA.
Therefore we likely oversampled people who have the resources and investment to engage with these communities, while neglecting those who do not.
On a distinct but related note, our response rates were low or nonexistent from important institutional types, in particular two year colleges, community colleges, and certain types of minority serving institutions.
We also limited our scope to lab instructors working in the US; we expect that our findings will not transfer directly to professional contexts in other countries.
 
More generally, our sample population necessitates caveats and caution when generalizing to lab instructors overall, and our results should not be interpreted as representative of the overall population of physics lab instructors. 
Rather, they represent a subset of that population that was subject to selection bias. 
Nevertheless, our findings illustrate important features of instructional lab professional contexts, and suggest some ways in which lab instructors are a distinct stakeholder group in the space of physics instructional labs. 
We claim that our results hold sufficient theoretical generalizability to suggest areas of focus in supporting and valuing lab instructors. 
We hope that this work guides further studies, in partnership with physics lab instructors, that recognize them as key stakeholders in instructional lab pedagogy.

\section{Summary and Conclusion}
Through surveys and interviews with individuals in physics departments across the US, we have started to paint a picture of the professional contexts of physics lab instructors.
Our analysis characterizes the people, institutional and organizational contexts, and resources involved in the jobs of physics lab instructors and supporters.
We have found successful lab instructors to be competent in a wide variety of technical, pedagogical, and interpersonal ways, as is required to succeed in a profession with wide ranging job descriptions that are often unofficial, unclear, and malleable.

While these characteristics can give lab instructors significant levels of personal agency, they also result in a sense of liminality or in-between-ness for lab instructors.
That sense is not necessarily a bad thing.
In fact, lab instructors often prided themselves on the ability to wear many hats, and by and large, their work and their instructional labs were valued in a variety of ways.
We suggest that it is this wide and overlapping scope of the physics lab instructor profession that makes lab instructors a unique, complex, and important population to study. 

We also suggest, as an avenue for further study, that the in-between-ness of lab instructor positions interacts with overall physics culture in complex ways.
One of the reasons lab instructor roles are so multifaceted and far-ranging is that lab work itself is unpredictable, complex, and uncertain; put simply, it is messy.
Such messiness stands at odds with the notion of physics's objectivity as a pure science that brings order and symmetry to the unwieldy complexity of nature.
It is possible that the predominant whiteness of physics lab instructors may afford many of us the privilege of championing such messiness in the lab while simultaneously claiming the (false) objectivity of physics's culture of no culture. 
Still, there are calls within laboratory education to make labs messier (in the sense of more authentic, uncertain, open-ended, or exploratory), which in a sense begin to break down the notion of physics's perfect objectivity.
Further research and further pedagogical reforms are needed to explore what such pedagogical movements could mean for lab instructors, and for physics culture more broadly.

In conclusion, by identifying what is needed to be successful in instructional lab professions, and by understanding the constraints that lab instructors operate within, the broader physics education community can better partner with, support, diversify, and elevate this critical population. 
Such efforts will in turn make experimental physics education better for students as well as for lab instructors and supporters.

\begin{acknowledgments}
We acknowledge the Advanced Laboratory Physics Association (ALPhA), and the American Association of Physics Teachers (AAPT) Committee on Labs, for their support in publicising and recruiting for this study.
We also gratefully acknowledge the students at WPI who assisted with database management in our recruiting, in particular Alice Gaehring, Bella Lenoce, and Robert Gyurcsan. 
Lastly, we wholeheartedly thank our survey respondents and our interviewees for their time and willingness to share their experiences and perspectives.
\end{acknowledgments}

\appendix
\newpage

\section{Survey items}
\textit{For review, survey items are appended as a PDF.}

\section{Interview protocol}
Instructional Labs Staffing Project: Interview Protocol
L Dana, B Pollard, S Mueller

We will send them the consent form electronically beforehand, and ask them to sign it electronically before the interview starts. That way, we can have Zoom record the meeting from the start, and we won’t need to take time during the interview to figure out electronic signatures.

We will also send them their answers to the survey as a PDF before the interview so they can refer to it when it comes up during the interview.
\subsection{Introduction}
So my name is Dana, I am the Lab Manager at Worcester Polytechnic Institute in Worcester, Massachusetts. In addition to that I am also a graduate student. 

The purpose of this interview is to learn about the structure, environment, and resources of the people that teach and run instructional physics labs. These interviews will investigate and document the landscape of a critical professional community in physics education. To that end, it’s really important for me to hear your perspective on your career to this point, your current job, and your place in your current institution. This is not an evaluation of you, your course, your institution, or your students; rather, I’m just exploring the various contexts that lab instructors find themselves in. Do you have any questions for me about the purpose of the interview, the consent form, or how your responses will be used?

This interview has four parts. First, I’ll ask you about your career path and current position. Then we’ll talk about the professional relationships you have in your work. Then I’ve got some questions about the resources you have available in your job. Lastly, I’ll finish with a few questions on how your work is valued in your institution. So if that sounds good to you, I'll go ahead and start by asking some questions about your career path and current position.

\subsection{Career path and org structure}
\begin{itemize}
    \item In the survey, you wrote that your job title is $<>$. What is your job description, in your own words? 
\item What are the qualifications for your position?
\item What was your career path that brought you to this job? 
\item Why did you apply for your position?
\item Is there a promotion path? 
\item In the survey, you wrote that you have been in your job for $<>$. How “permanent” is your position?
\end{itemize}

\subsection{Relationships and identities}
\begin{itemize}
    \item In the survey, you wrote that your boss is $<>$, and $<>$ report to you. You also mentioned that you work with $<>$ in your department, $<>$ at your institution, and $<>$ outside your institution. Did I get all that correct?
    \item Of all of those relationships you talked about above, which are the most meaningful/valuable/impactful?
\item In what way is your relationship with these other people meaningful? 
\item How do your personal identities affect your interactions with these people? – Specific followup if someone is struggling – Some examples of personal identities might be gender, faculty vs staff, race, field, disability status etc. 
\item How much control do you have over your day-to-day schedule and how you spend your time? Task driven vs schedule driven 

\end{itemize}
 
\subsection{Resources and agency}
\begin{itemize}
    \item You indicated in the survey that you have access to this equipment: $<>$.  Is there anything to add to that list?
\item What is the funding situation available to you to buy/update equipment?
\item Do you think your lab is well funded? In need of more funding? 
Why do you think that number was chosen? 
\item What are some of the items you have to budget for or not buy in order to make do? What is an item you would like to buy but could not due to budget issues? 
\item Is there anything on your wish list you would have loved to buy but it’s not in the budget
\item You indicated in the survey that $<$you do or do not interact with student employees$>$. Did I get that right, and is there anything to add?
\item How many students do you hire in a typical semester? 
\item What do you look for in a student employee. 

\end{itemize}
 For this section, I’ll be referring to “your labs,” so first I’d like to talk a bit about what that phrase means for you.
 \begin{itemize}
     \item Which lab courses or lab spaces do you consider “yours?”
\item What makes you feel that those courses are yours?
\item What is it about those courses that make them yours?
\end{itemize}

Next I have a series of questions about how much control or agency you have over various aspects of your lab(s). As we go through, feel free to mention other people who have control with or instead of you, and if the timescale matters (for example, day-to-day versus semester-to-semester).
\begin{itemize}
    \item How much control do you have in determining the learning goals of your labs? Course vs individual labs? 
\item How much control do you have in determining the experiments, supplies, and apparatus done in your labs, and how and where they are used?
\item How much control do you have over the lab guides and/or written documentation that are used in your labs?
\item How much control do you have over the course structure, scheduling, grading, and logistics of your labs?
\item How much control do you have in hiring students and/or support people for your labs?
\item What do students think of working in your labs? 
\item What is your management style in managing those students? 
\item What do you think helped most in driving your vision for the lab reform? 
\end{itemize}

\subsection{Value from institution} 
\begin{itemize}
\item You ranked $<>$ in the survey about how much instructional labs are valued (or not) by the institution. Please describe a time when you felt that instructional labs were (not) valued by your institution.

\item I’d now like to ask a similar question about you in particular. You ranked $<>$ in the survey about how much you are valued (or not) by the institution. Please describe a time when you felt that your work was (not) valued by your institution.
Or past institutions?
\item Is there a difference between upper admin and interdepartmental treatment? 
\end{itemize}

\subsection{Wrap-up}
Thank you so much! Those were all the specific questions I had.
Is there anything else you'd like to tell me?
Thank you very much for your help with this project! My colleagues and I will be analyzing these interviews over the next few months. We plan on writing up our findings for peer review, and presenting them at the next ALPhA meeting. Before those are finalized, there might be opportunities for you to review the trends that we find and provide additional information. We will be in touch about that as things progress.

\bibliography{pollardRefs}
\end{document}